\input harvmac

\def\p{\partial}
\def\ap{\alpha'}

\Title{EFI-96-36}{Open Membranes in Matrix Theory}
\centerline{Miao Li}
\centerline{\it Enrico Fermi Institute}
\centerline{\it University of Chicago}
\centerline{\it 5640 Ellis Avenue, Chicago, IL 60637} 
\centerline{\tt mli@curie.uchicago.edu}

\bigskip
\centerline{\it }
\centerline{\it }
\centerline{\it } 
\bigskip

We discuss how to construct open membranes in the recently proposed
matrix model of M theory. In order to sustain an open membrane, two
boundary terms are needed in the construction. These boundary terms 
are available in the system of the longitudinal five-branes and D0-branes.

\Date{December 1996} 

\nref\bfss{T. Banks, W. Fischler, S. H. Shenker and L. Susskind,
hep-th/9610043.}
\nref\dhn{B. de Wit, J. Hoppe and H. Nicolai, Nucl. Phys. B305 [FS 23]
(1988) 545; B. de Wit, M. Luscher and  H. Nicolai, Nucl. Phys.
B320 (1989) 135.} 
\nref\bd{M. Berkooz and M. R. Douglas, hep-th/9610236.}
\nref\sgrt{L. Susskind, hep-th/9611164; O. Ganor, S. Ramgoolam and
W. Taylor IV, hep-th/9611202.}
\nref\ab{O. Aharony and M. Berkooz, hep-th/9611215.}
\nref\mike{M. R. Douglas, hep-th/9612126.}
\nref\st{A. Strominger, hep-th/9512059; P. K. Townsend, hep-th/9512062.}
\nref\dkps{M. R. Douglas, D. Kabat, P. Pouliot and S. H. Shenker,
hep-th/9608024.}
M theory is recently conjectured to be described by a large N quantum
mechanical system based on D-particle dynamics \bfss. One of the major clues
for this conjecture is the realization of the light-cone membrane
Hamiltonian as a large N gauged quantum mechanics \dhn. The conjecture
is further supported by an explicit calculation of the membrane
tension \bfss\ and by the Dirac quantization of the membrane charge
in the background of a longitudinal five-brane \bd. Further supporting
evidence has been collected in \sgrt\  \ab\ \mike.

A closed, transverse  membrane with toroidal topology can be constructed 
by utilizing the basis for large N matrices spanned by $U, V$ with 
$UV=\exp (2\pi i/N)VU$. $U$ and $V$ in turn can be realized with
$U=\exp(ip)$, $V=\exp(iq)$, $[q,p]=2\pi i/N$. Now $p$ and $q$ parametrize
a torus of unit radii, it is then natural that only closed membranes
are realized using this basis. We wish to point out in this note
that with a slight abuse, this basis can be used to construct open,
transverse membranes with cylindric topology. Before proceeding to
our construction, it is useful to briefly review the light-cone
Hamiltonian of \bfss\ and the construction of closed membranes there. 

Compactifying M theory on a circle of radius $R$, D0-branes emerge
as supergravitons carrying positive unit momentum $p_{11}=1/R$. These
particles, regarded as massive particles in 10 dimensions, has a 
mass $m=p_{11}$. The light-cone Hamiltonian of multiple particles
is given by the dimensionally reduced 10D super Yang-Mills theory.
In the temporal gauge $A_0=0$, the bosonic part of the Hamiltonian reads

\eqn\ham{H=P_-={m\over 2}\Tr\left (\sum_i (\dot{X}^i)^2-C
\sum_{i<j}[X^i,X^j]^2\right),}
where constant $C=(2\pi\ap)^{-2}$. A number of properties of M theory
have been checked using this Hamiltonian, as we mentioned before.
In particular, a transverse membrane is constructed by $X^8=R_8p$,
$X^9=R_9q$ and $X^i=0$, $i=1,\dots, 7$. Here $p$ and $q$ both have
a period $2\pi$, so $X^8$ and $X^9$ have periods $2\pi R_8$ and 
$2\pi R_9$ respectively. Note that the configuration satisfies
the stationary equations of motion
$$\sum_i[X^i,[X^j,X^i]]=0$$
and $[X^8,X^9]\ne 0$. This is possible only in the large N limit,
since for a finite N, the equations of motion always imply 
$[X^i,X^j]=0$. The light-cone Hamiltonian is not vanishing, and
the mass squared of the membrane is given by the relation $M^2=
2P_{11}H=2NmH$. This is shown to agree with the membrane tension 
formula \bfss.

It appears that it is impossible to discuss open membranes with
the basis $(p,q)$, as it has toroidal topology. One way to get
around this problem is to introduce a pair of noncommutative
variables parametrizing a finite cylinder. We don't know how to
do this at this moment. Instead we propose to study open membranes
using the basis $(p,q)$ with a slight modification for the ansatz $X^8$, $X^9$.
Since it is generally believed that only wrapped membrane is stable,
we shall still compactify $X^9$ with radius $R_9$, and leave $X^8$
uncompactified. The open membrane we are interested in will wrap around
$X^9$ once, and stretched from $X^8=x_1$ to $X^8=x_2$. The ansatz
we propose is the following

\eqn\ansatz{X^9=R_9p,\quad 
X^8=\left\{  \matrix{   x_1, & q<q_1\cr
x_2, &q>q_2\cr
{x_2-x_1\over q_2-q_1}q+{x_1q_2-x_2q_1\over q_2-q_1}, &q_1<q<q_2}\right\}
}
where $q_1\ge 0$ and $q_2\le 2\pi$. The strategy here is to break
the circle parametrized by $q$ by assuming that $X^8$ collapses to a 
constant point at both ends.
It is important to keep in mind that $X^8$ is a function of
only $q$, and is diagonal in the diagonal basis of $q$. Let $X^8=f(q)$.
Taking derivative twice, 
\eqn\delt{f^{(2)}(q)={\Delta x\over \Delta q}\left(\delta (q-q_1)-
\delta (q-q_2)\right).}
Before we proceed to justify the above equation, we compute the membrane
tension first. The commutator $[X^8,X^9]=R_9(2\pi i/N)f'(q)$, and the
first derivative of $f(q)$ is a step function. It is straightforward
to evaluate the light-cone Hamiltonian, using the method presented
in \bfss. The mass squared of the open membrane is 
$$M^2={m^2(2\pi R_9\Delta x)^2\over 2\pi\Delta q} ,$$
where we take $2\pi\ap =1$.
This gives rise to a membrane tension 

\eqn\lowb{T_2^2={m^2\over 2\pi\Delta q}\ge ({m\over 2\pi })^2,}
since $\Delta q\le 2\pi$. We see that the membrane tension is bounded
from below by the true membrane tension. With $\Delta q< 2\pi$,
the open membrane must be interpreted as meta-stable, since its spectrum
is continuous, and nothing prevents it from decaying to its lower bound.
Actually, it will be shown later that only in the limit $\Delta q=2\pi$,
it is possible to maintain some unbroken supersymmetry.

Next, we show that \delt\ can be interpreted as forces needed to sustain
the open membrane at its boundaries. The two relevant equations of 
motion resulting from ansatz \ansatz\ are
\eqn\eom{\eqalign{ [X^8,[X^9,X^8]]&=0,\cr
[X^9,[X^8,X^9]]&=({2\pi R_9\over N})^2{\Delta x\over \Delta q}\left( 
\delta(q-q_1)-\delta(q-q_2)\right).}}
Roughly speaking, the second equation implies that a force is needed
at boundary $X^8=x_1$ and an opposite force is needed at boundary
$X^8=x_2$, in order to sustain the open membrane. 

The physical origin of such force must be found in the system consisting
of five-branes and D0-branes.  Such a system is discussed in \bd.
A transverse open membrane can end
on a longitudinal five-brane, and the latter is interpreted as a 
D4-brane in 10 dimensions. 
Typically, what has been considered before in the M theory context
is a static open membrane stretched between two parallel five-branes \st.
Due to 11 dimensional Lorentz invariance, one can always boost the
open membrane in one of the longitudinal directions of the five-branes.
Here in the matrix model context, a transverse membrane always carries
an infinite longitudinal momentum, that is, it moves in the speed of
light along $X^{11}$. To see that the boundary terms in \eom\ are actually
available  in the system studied in \bd, we first find out the
corresponding term in the Hamiltonian phenomenologically. With a
finite N, $q$ has eigen-values $q=2\pi (i-1)/N$, $i=1,\dots, N$.
Let the $i$'s corresponding to $q_1$ and $q_2$ be $i_1$ and $i_2$. The
delta function $\delta (q-q_1)$ can be replaced by a diagonal matrix
(in the basis in which $q$ is diagonal) with only one nonvanishing 
entry at the $i_1$-th row and the $i_1$-th column with a value
$N/(2\pi)$. Let $v_i$ be the unit vector with only one nonvanishing entry
at the $i$-th row, then $\delta(q-q_1)=N/(2\pi) v_{i_1}v_{i_1}^+$.
A term in the Hamiltonian such as
\eqn\add{\Delta H=mR^2_9({2\pi\over N})({\Delta x\over\Delta q})\Tr X^8
\left(v_{i_1}v_{i_1}^+ -v_{i_2}v_{i_2}^+\right)}
would reproduce the desired boundary terms in \eom.

Now, consider two parallel longitudinal five-branes, with 5 longitudinal
spatial coordinates $X^m$, and one of them is $X^{11}$, and 5 transverse
coordinates $X^a$. Let one of the longitudinal coordinates be $X^9$,
the same as the circle around which the open membrane is wrapped. And
let one of the transverse coordinates be $X^8$, along which the open
membrane is stretched. Further, let the location of the two five-branes
be $X^8=x_1, x_2$. In the background of these five-branes, there are
additional modes in the dynamics of D0-branes, these correspond to 
open strings stretched between a five-brane and a D0-brane. In particular,
there are two sets of bosons transforming in the fundamental representation of
$U(N)$, call them $V_I^{\rho}$, where $\rho$  index the spinor 
representation of the positive chirality
of $SO(4)_L$, the rotation group
of the 4 longitudinal directions excluding $X^{11}$, and $I=1,2$. We refer
to ref.\bd\ for further notation. The bosonic part of the additional 
Hamiltonian is
$$\Delta H=|\p_tV_I^{\rho}|^2+V_{I\rho}(X^a-x^a_I)^2
V_I^{\rho}-V_{I\rho}[X^m,X^n]\sigma^{\rho\sigma}_{mn}
V_{I\sigma}+|V_I|^4.$$
Since there is only one nontrivial $X^m$, that is $X^9=R_9p$, the third
term in the above Hamiltonian drops out, hence the equation of
motion for $X^9$ is just $[X^8,[X^9,X^8]]=0$ which is satisfied by
our ansatz. So the only nontrivial terms relevant to our problem are
\eqn\addm{\Delta H=|\p_tV_I^{\rho}|^2+V_{I\rho}
(X^8-x^8_I)^2V_I^{\rho} +|V_I|^4,}
where $x^8_1=x_1$, $x^8_2=x_2$. This additional part is
very similar to the required term in \add. Indeed, the above Hamiltonian
will reproduce the desired boundary terms provided
\eqn\cond{\eqalign{ 2(X^8-x_1)V_1V_1^+ &=mR_9^2({2\pi\over N})
({\Delta x\over \Delta q})v_{i_1}v_{i_1}^+, \cr
2(X^8-x_2)V_2V_2^+ &=-mR_9^2({2\pi\over N})
({\Delta x\over \Delta q})v_{i_2}v_{i_2}^+.}}
So in order to have the right boundary terms, the vector fields $V_I$
must be excited at the boundary, or the vector fields must have
nonvanishing quantum fluctuations close to the boundary. Before exploring
this possibility, let us first notice that the above equations can be
satisfied in principle. Examine the first equation. The R.H.S. tells
us that $V_1$ must have only one nonvanishing entry at the $i_1$-th 
row. Since all eigenvalues of $X^8$ must be equal to or greater than
$x_1$, so $X^8-x_1\ge 0$, and the signs on the both sides agree. In 
the second equation, $X^8-x_2\le 0$, again the signs on the both sides
of the second equation agree.

Actually, the problem is
a little more involved than represented by \cond. The fermions transforming
as vector of $U(N)$ will likely to contribute to the boundary forces.
We believe that the physics is however captured by \cond.
The difficulty to satisfy \cond\ at the classical level is that the
potential for the vectors $V_I$ is always positive, so it is impossible
to give $V_I$ a vacuum expectation value, even only at boundaries of
the open membrane. Quantum mechanically, the possibility for the existence
of nonvanishing $V_I$ at boundaries exists. 
Consider a single D0-brane
close to one of the five-brane. It is known there is an attractive force
between the two objects, and a bound state can form \dkps. The typical
transverse radius for such a bound state is just the 11 dimensional
Planck length. Such a force 
is not known if there are many D0-branes and they are not arranged
in a commutative fashion. 

Eqs.\cond\ imply that $V_1\sim v_{i_1}$ and
$V_2\sim v_{i_2}$, so $X^8-x_I$ vanishes when applied to $V_IV_I^+$.
This is due to the fact that we have identified $x_I$ in \ansatz\ with $x_I$
in \addm. Physically one may introduce a cut-off for $X^8-x_I$. 
This cut-off distance
between a boundary of the open membrane and a five-brane is constrained
by the condition that the additional Hamiltonian \addm\ should not
contribute to the membrane tension in the large N limit. Let, say
$(X^8-x_1)$ in \cond\ scales as $1/N^\alpha$ in the large N limit,
then $|V_1|^2\sim 1/N^{1-\alpha}$ according to \cond. The second term
in \addm\ will contribute an amount $1/N^{1+\alpha}$ to the Hamiltonian.
The condition that this term can be neglected in computing the membrane
tension is $\alpha>0$. (This implies that the cut-off $1/N^\alpha
\rightarrow 0$ in the large N limit.) On the other hand, the last term 
in \addm\ scales as $1/N^{2(1-\alpha )}$ and can be neglected if 
$\alpha <1/2$. 

We need to
examine the supersymmetry transformation in order to see which 
value of the parameter $\Delta q$ is allowed. It can be seen that
supersymmetry vanishes acting on all fields except $\theta$, the
superpartner of $X^\mu$. The SUSY transformation of this field,
using the notation of \bd, is
\eqn\susy{\delta\theta^{\rho}_\alpha =2[X^8,X^9](\gamma_{89}
\eta)^{\rho}_\alpha 
+\eta'^{\rho}_\alpha  +V^{(\rho}_IV^{\sigma)}_{I}
\eta_{\sigma\alpha},}
where the new index $\alpha$ is the spinor index of the transverse
rotation group $SO(5)$.
For a closed membrane considered in \bfss, there is no a third term.
The first term is proportional to the identity matrix, so can be cancelled
by the second term. For the open membrane, the first term is not
proportional to the identity matrix, a step function sets in.
The commutator is
$$[X^8,X^9]={2\pi i\over N}R_9{\Delta x\over \Delta q}\left(
\theta(q-q_1)-\theta(q-q_2)\right),$$
and vanishes when $q<q_1$ or $q>q_2$. As in the closed membrane case,
one can always choose $\eta'$ to cancel the first term for $q_1<q<q_2$.
For $q$ outside this range, the first term vanishes, then the constant
term $\eta'$ must be cancelled by the third term. However, in the
large N limit, $\eta'$ has
infinitely many nonvanishing diagonal entries to be cancelled by the third 
term if, say,
$q_1>0$. This can not be constructed from a finite number vectors
$V_1^{\rho\dot{\rho}}$, since the third term will be close to a 
``pure state'' density matrix, while the part of $\eta'$ to be
cancelled is rather a ``mixed state'' density matrix. A resolution
of this problem again comes from the fact that $|V|^2$ is nonvanishing
only at the quantum level. If one interprets $V_IV_I^+$
as the quantum average $\langle V_IV_I^+ \rangle$, then this matrix
can be a ``mixed state'' density matrix. From cancelation of SUSY
\susy, we deduce that 
\eqn\sur{\langle V_1V_1^+ \rangle\sim {1\over N}\sum_{i=1}^{i_1}v_iv_i^+ ,}
and a similar result for $\langle V_2V_2^+ \rangle$.

The above ansatz contradicts \cond\ however, since there only a single
$v_{i_1}$ appears. To resolve this problem, recall that there is
certain arbitrariness in regularizing the delta functions in \eom.
If $q_1\sim i_1/N \rightarrow 0$ in the large N limit (we shall show 
this is the case indeed), then it is equally good to use the following
$$\delta (q-q_1)={N\over 2\pi i_1}\sum_{i=1}^{i_1}v_iv_i^+ ,$$
and consequently the R.H.S. of \cond\ is modified. Use \sur\
in this modified condition, we find $i_1\sim N^\alpha$. The condition
that the second term of \addm\ does not contribute to the membrane
tension is still $\alpha>0$. The condition that the third term
does not contribute to the membrane tension is $|V_1|^4\sim
i^2_1/N^2<< 1/N$, so $\alpha <1/2$, also the same as the condition we derived
using a single $v_{i_1}$. Finally, we see that 
\eqn\lim{q_1=i_1/N\sim 1/N^{1-\alpha} \rightarrow 0.}
Similarly, in the large N limit $q_2\rightarrow 2\pi$, and $\Delta q
\rightarrow 2\pi$. It must be emphasized that although $q_1\rightarrow
0$, but $i_1\sim N^\alpha\rightarrow \infty$, and this justifies
our ansatz \ansatz\ and equations of motion \eom.

It remains to determine the exponent $\alpha$. A closer examination of the
large N dynamics is necessary in order to do this and to work out more 
details of our construction. Here we shall make only a plausible
guess for the boundary dynamics. In general, one may replace our
ansatz with a general boundary function such that equations of motion
\eom\ are still valid at the boundary. With SUSY \susy\ unbroken for
some choice of $\eta$ and $\eta'$, generally one expects that the one-loop
correction is vanishing. In particular, the large one-loop kinetic
energy in \addm\ must be canceled by that of fermions. It is plausible
that the major contribution to the second and third terms in \addm\ is
from higher loops. If so, in the perturbative calculation, the third term
becomes important. If one assumes that these two terms are comparable,
then one deduces $i_1\sim N^\alpha=N^{1/3}$, that is, the exponent
$\alpha=1/3$.

Finally, a remark on central charges. A closed membrane carries a rank
2 tensor central charge. In an appropriate formulation of super algebra
of the matrix model, this central charge should be of form $\tr[X^i,X^j]$.
The closed membrane solution indeed has a nonvanishing commutator. For an 
open membrane, its ends appear as a closed string in the five-branes.
A closed string in a five-brane carries a central charge corresponding
to the anti-self-dual tensor field in the tensor multiplet. Again,
in an appropriate formulation of super algebra, one expects a vector central
charge. In the matrix model, the natural candidate for this is just
$\sum_j [X^j,[X^i,X^j]]$. This quantity must have nonvanishing value
only at a boundary of the open membrane. This further justifies our
equations of motion \eom.

\noindent{\bf Acknowledgments} 

We would like to thank Mike Douglas for comments. We are informed by him
that a similar result has been obtained by Berkooz and Douglas. We
wish to thank Emil Martinec for a discussion on central charges.
This work was supported by DOE grant DE-FG02-90ER-40560 and NSF grant
PHY 91-23780.

\listrefs

\end